# Impact of COVID-19 behavioral inertia on reopening strategies for New York City Transit

**Ding Wang, Brian Yueshuai He, Jingqin Gao, Joseph Y. J. Chow\*, Kaan Ozbay, Shri Iyer**
C2SMART University Transportation Center, New York University Tandon School of Engineering, Brooklyn, NY, USA
\*Corresponding author email: joseph.chow@nyu.edu

**Abstract**

The COVID-19 pandemic has affected travel behaviors and transportation system operations, and cities are grappling with what policies can be effective for a phased reopening shaped by social distancing. A baseline model was previously developed and calibrated for pre-COVID conditions as MATSim-NYC. A new COVID model is calibrated that represents travel behavior during the COVID-19 pandemic by recalibrating the population agendas to include work-from-home and re-estimating the mode choice model for MATSim-NYC to fit observed traffic and transit ridership data. Assuming the change in behavior exhibits inertia during reopening, we analyze the increase in car traffic due to the phased reopen plan guided by the state government of New York. Four reopening phases and two reopening scenarios (with and without transit capacity restrictions) are analyzed. A Phase 4 reopening with 100% transit capacity may only see as much as 73% of pre-COVID ridership and an increase in the number of car trips by as much as 142% of pre-pandemic levels. Limiting transit capacity to 50% would decrease transit ridership further from 73% to 64% while increasing car trips to as much as 143% of pre-pandemic levels. While the increase appears small, the impact on consumer surplus is disproportionately large due to already increased traffic congestion. Many of the trips also get shifted to other modes like micromobility. The findings imply that a transit capacity restriction policy during reopening needs to be accompanied by (1) support for micromobility modes, particularly in non-Manhattan boroughs, and (2) congestion alleviation policies that focus on reducing traffic in Manhattan, such as cordon-based pricing.

## 1. Introduction

When a pandemic spread to a large population, response strategies to "flatten the curve" of the spread may require parts of society to "stay at home" and to maintain social distancing (Thunström et al., 2020; Cohen and Kupferschmidt, 2020). In the case of COVID-19, the "stay-at-home" order and implications of social distancing measures have shut down many public services, retail businesses and tourism activities. These have profoundly impacted the economic activities in many cities around the world which in turn have deeply changed people's travel behavior. For example, data reveal that people have become more afraid to take public transit or other shared modes and avoid gathering in crowded areas (Pakpour and Griffiths, 2020). As a community reopens, the changed behavior may exhibit *inertia* (Ghader et al., 2020; Shadmehr and de Mesquita, 2020; Liu et al., 2017; Alós-Ferrer et al., 2016; Cherchi et al., 2014; Xie et al., 2014; Srinivasan and Mahmassani, 2000; Chorus and Dellaert, 2009), where travelers continue to behave according to the stay-at-home setting due to residual fears.

In addition to behavior, the spread of COVID-19 has raised new challenges for public transportation systems (Bóta et al., 2017; Hajdu et al., 2019; El Shoghri et al., 2019; Qian et al., 2020; Mo et al., 2020; Shoghri et al., 2020), especially for transit agencies in New York City (NYC), as it is one of the hardest-hit cities in the world. Many of these challenges deal with how the pandemic might spread through the transit systems, although empirical studies find little evidence of tracing outbreaks to transit services (Sadik-Khan and Solomonow, 2020).

With the need for social distancing and the added residual fears, another challenge exists in the case of reopening. As public transport is designed to mobilize people through shared usage and benefits dense populations the most, the pandemic has presented an existential crisis to such systems. Some cities are implementing strategies that limit the number of passengers on vehicles to keep riders and workers safe. For instance, the subway system in Beijing, China, limits the subway occupancy below 50% of its maximum capacity around February 2020 (XinhuaNet, 2020). In New Jersey (NJ), NJ Transit trains and buses was ordered to cut capacity to 50% until July 15, 2020 to maintain social distancing (NJ TRANSIT, 2020).

Policymakers need to carefully balance between encouraging dense operations and social distancing practices especially as sectors of the economy are reopened considering traveler behavioral inertia. Research questions arising from this pandemic include:
- How does the pandemic impact people's travel behavior during the pandemic?
- If there is indeed inertia in the impacted travel behavior, how would it play a role in the public transportation operations and road traffic in a re-opened state?
- How should public transportation systems, conventionally designed to encourage densification, operate in a reopened state that requires de-densification of the public?

With a shift away from public transit, there is a key concern that a bigger burden will be placed upon road traffic when cities reopen. A travel demand model is needed to answer these questions for NYC. As the "stay-at-home" order requires different proportions of industries to work from home (WFH), such a model needs to be sensitive to employment industries of commuters during the COVID-19 period. In addition, the model needs to be sensitive to different transit schedules and capacities as it changes from pre-pandemic through different reopening stages of COVID-19 (Burke, 2020).

Agent-based simulation models are effective in capturing the interactions between agents and transportation system to output the equilibrated simulation results at the agent level. Hackl and Dubernet (2019) developed a disease propagation model in MATSim to simulate epidemic outbreaks. The results show that even with simple assumptions, the agent-based model could give a good approximation compared to the observed data. However, as far as we know, no research in the literature have used the agent-based traffic simulation model to study the impact of the epidemic/pandemic disease on people's travel behaviors and mode share as well as its ability to test different transport system operating policies for decision-makers.

A synthetic population was developed for NYC (He et al., 2020a, Chow et al., 2020) for the 8M+ population in NYC that includes NAICS (North American Industry Classification System[1]) employment industries. Based on the synthetic population along with calibrated transit schedules, a multi-agent simulation (Horni et al., 2016) model, MATSim-NYC (He et al., 2020b, Chow et al., 2020) was developed by C2SMART researchers for NYC (Figure 1).

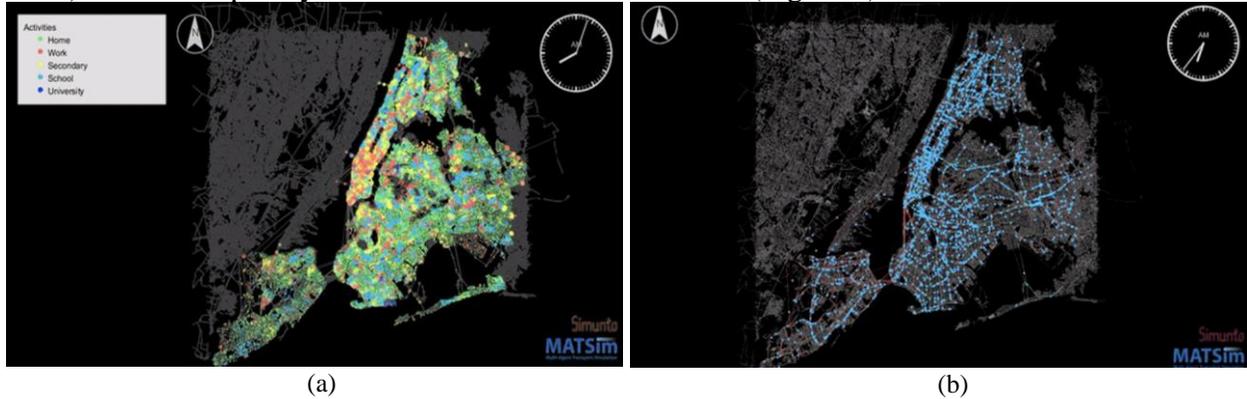

Figure 1. Illustration of (a) activity simulation and (b) mobility simulation in MATSim-NYC.

We propose to re-calibrate the MATSim-NYC model using ridership and WFH data during the COVID-19 stay-at-home period to update the mode choice utility functions for the synthetic population. The WFH data is obtained for different employment industries from a study by Dingel and Neiman (2020). The recalibrated model (which we call the MATSim-NYC-COVID model) can capture the effect of COVID-19 on shifts in mode preferences from users. Using this recalibrated model, we then compare the re-opening scenarios where the population is assumed to continue to exhibit the same mode preferences. Two reopening scenarios (with and without transit capacity restrictions) are analyzed to forecast the impacts on transit mode share and road traffic.

The unique contributions of this study include: (1) evaluating the sensitivity of road traffic to WFH policy for different employment industries; (2) insights to the mode choice behavioral effect of COVID-19 on NYC residents; (3) quantifying the impacts of the mode choice inertia on transit ridership and road traffic; and (4) evaluating the impact of a 50% capacity restriction on transit ridership. The developed model will also be useful for comparing different reopening plans.

The remainder of the study is organized as follows. Section 2 provides an overview of the data used, the existing MATSim-NYC model, and scenarios defined for this study. Section 3 describes the recalibrated model under COVID-19 and its validation and analysis. Section 4 presents the analysis of different transit capacity restriction under different reopening phases as proposed by New York State (New York State, 2020).

---

[1]North American Industry Classification System (NAICS). https://www.bls.gov/bls/naics.htm

## 2. Data and scenarios

This section provides the synthetic population data and MATSim model data used for building the simulation model, as well as the relevant details of the MATSim-NYC model. MTA transit ridership data (MTA, 2020) and Apple mobility trends report data (Apple, 2020) are used to calibrate and validate the COVID model. We also introduce the existing MATSim-NYC model and the scenarios that will be tested in this study.

### 2.1. Data
*Synthetic Population data and MATSim model data*
The American Community Survey, 2016 Longitudinal Employer-Household Dynamics (LEHD), and 2040 Socioeconomic and Demographic Forecasts (SED) were used to generate personal and household attributes of the synthetic population (Chow et al., 2020). The 2010/2011 Regional Household Travel Survey (RHTS) data was employed to prepare travel agendas and model mode choice. To incorporate emerging modes (bike-sharing and ride-hailing), 2016 trip count data of Citi Bike and For-Hire-Vehicles (FHV) were also adopted. The 2017 Citywide Mobility Survey data was used to validate the city-level mode share of the synthetic population.

The input network was developed with a road network transformed from OpenStreetMap (OSM) data and a transit network and schedule generated from General Transit Feed Specification (GTFS) data. INRIX speed data and the 2016 New York City Bridge Traffic Volumes data were used to calibrate the road network's link speeds and capacities.

*MTA Turnstile Data and Apple Mobility Trends Report*
Metropolitan Transportation Authority (MTA), the largest public transit authority in the US, typically carries over 11 million transit riders on an average weekday. Its bridges and tunnels serve more than 800,000 vehicles each weekday and carry more traffic than any other bridge and tunnel authority in the nation. The implementation of the stay-at-home order on March 2020 has an immediate impact on the transit ridership and traffic volume in NYC. During COVID-19, both subway ridership and vehicular traffic on MTA facilities show steep declines. The decline rates reached up to 92% in peak transit ridership as the week of April 6, 2020 to April 12, 2020, and up to 69% in vehicle traffic through MTA bridges and tunnels at the same time, compared to the same date/week in 2019.

This data was supplemented by the Apple Mobility Trends report that reflects requests for directions in Apple Maps. According to the daily data posted on April 18, 2020, the trips by transit, driving and walking have decreased 88%, 54%, 76%, respectively. Table 1 shows the weekly changes of the traffic data during the COVID period according to the MTA data and Apple mobility trends report. The average trip reduction data from March 23, 2020 to April 19, 2020 was used to calibrate the COVID model.

### 2.2. MATSim-NYC
MATSim is an open-source framework for implementing large-scale agent-based traffic simulations (Horni et al., 2016). The MATSim platform can simulate a large-scale transport on a per-agent timestep-based level. The model is initialized with agents in the synthetic population and each agent has an initial plan during the simulated day. Those plans consist two elements: one is a set of activities, where each activity has a location with start/end time, e.g. "home" and "work" are two common activities that usually exist in agent's plan. The second element is the "Leg" which provides the connections between two activities via a mode of transport (e.g. "car", "transit", "walk"), the route taken, etc.

Table 1 Weekly changes of traffic data

| Date | MTA data | | Apple mobility trends | | |
|---|---|---|---|---|---|
| | **Subway Ridership** | **Vehicle Traffic via MTA bridges and Tunnels** | **Driving** | **Transit** | **Walking** |
| 3/23/2020 - 3/29/2020 | -87.00% | -64.00% | 60.34% | 84.75% | 76.48% |
| 3/30/2020 - 4/5/2020 | -91% | -68% | 59.91% | 86.18% | 76.07% |
| 4/6/2020 - 4/12/2020 | -92% | -69% | 58.83% | 87.00% | 75.52% |
| 4/13/2020 - 4/19/2020 | -85% | -66% | 54.80% | 86.88% | 75.04% |
| Average | -88.75% | -66.75% | 58.47% | 86.21% | 75.78% |

There are three components in MATSim: execution, re-planning and scoring (Balmer, 2007; Horni et al., 2016). The objective of MATSim is to optimize the daily plans for each agent by iteratively running the three components day by day (e.g., 100 iterations indicate 100 days of updated daily plans). In the execution module, all agents choose one plan and execute their chosen plan. The scoring module uses a utility function to evaluate the performance of each agent's plan in the execution module. The re-planning module adjusts the plan elements (e.g., departure time, traffic mode) according to the plan score and adapts the plans to the traffic condition. Figure 2 shows the modeling framework used in MATSim.

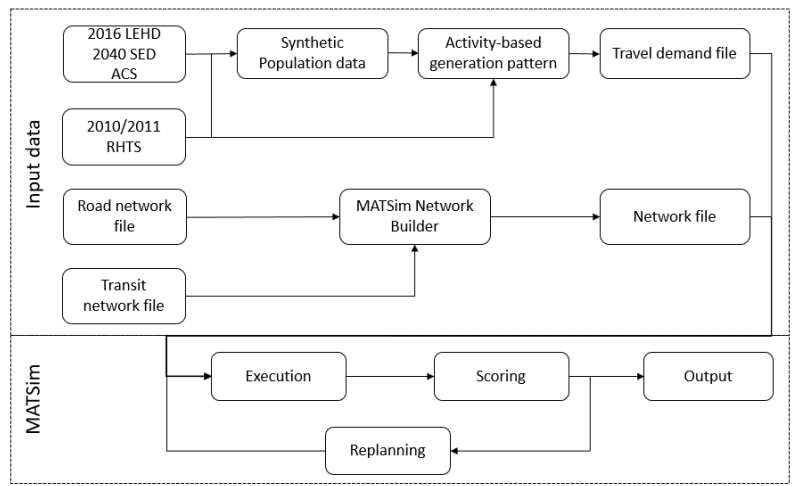

Figure 2. Modeling Framework.

A synthetic population of NYC was created which incorporates the demographic information and travel patterns of 8.24 million people for the base year of 2016 of the city. This results in 30,991,820 average daily trips made by the synthetic population. For this synthetic population, a tour-based nested logit model was estimated for Manhattan and non-Manhattan population segments.

The modes Driving Alone, Carpool, Public Transit, Taxi, Bike, and Walk were estimated from the Regional Household Travel Survey. For-Hire Vehicles (FHV) and Citi Bike were calibrated to have cost and travel time coefficients similar to Taxi and Bike, respectively, with the alternative specific constants (ASCs) fitted to have the model outputs match the corresponding trip count data in 2016. Those modes were further estimated using the choice of a separately estimated smartphone ownership model as a feature for Citi Bike and as a condition for alternative availability for FHV. The validation of the model was conducted using the 2017 Citywide Mobility Survey provided by New York City Department of Transportation (NYCDOT) and shown to be a good fit.

Since MATSim's mode utility functions are assumed to follow a flat multinomial logit (MNL) structure as opposed to nested structures, the estimated mode choice model was converted into an equivalent trip-based MNL structure. The parameters (alternative specific constant and cost) of the driving nest were multiplied by the nest scale factor and adopted as the approximate parameters in the equivalent trip-based MNL model (He et al., 2020; Chow et al., 2020). The travel time parameter of driving mode was set to be zero according to the MATSim guidebook (Nagel et al., 2016). Mode choice utility function parameters (Chow et al., 2020) are shown in Table 2. These values represent the pre-COVID mode choice behavior.

Table 2 MATSim-NYC mode choice utility function parameters (source: Chow et al., 2020)

| Manhattan | | car | carpool | transit | taxi | bike | walk | Citi Bike | FHV |
|---|---|---|---|---|---|---|---|---|---|
| Constant | | -0.06 | 0.00 | 2.95 | 1.06 | 0.44 | 5.73 | -0.37 | 0.79 |
| Travel Time | | 0 | 2.35 | 0.00 | 1.75 | -2.55 | -3.94 | -2.55 | 1.75 |
| Cost | | -0.06 | | | | | | | |
| Transit | Access Time | -0.96 | | | | | | | |
| | Egress Time | -0.86 | | | | | | | |
| | Transfer Time | -1.46 | | | | | | | |
| Non-Manhattan | | car | carpool | transit | taxi | bike | walk | Citi Bike | FHV |
| Constant | | -0.05 | 0.00 | 0.76 | -1.81 | -1.35 | 3.49 | -2.04 | -3.38 |
| Time | | 0.00 | 0.36 | 0.00 | 0.00 | -5.64 | -5.05 | -5.64 | 0.00 |
| Cost | | 0 | | | | | | | |
| Transit | Access Time | -1.71 | | | | | | | |
| | Egress Time | -1.67 | | | | | | | |
| | Transfer Time | -1.61 | | | | | | | |

For the network topology, the base topology was converted into a network in MATSim from Open Street Map (OSM) and a transit network generated from General Transit Feed Specification (GTFS) data. For computation efficiency, the population in the simulation is scaled to 4% of the real population (by comparison, MATSim models in other cities like Zurich typically use a 10% scaled population). The road network is shown in Figure 3(left), while the green layer in Figure 3 (right) shows the transit network.

INRIX data were used to calibrate unsaturated road speeds while the bridge crossing average annual daily traffic (AADT) data were used to calibrate the road capacities (see Chow et al., 2020). The validation of the MATSim-NYC trip assignment was conducted by comparing the outputs to two data sets: ten stations from the 2016 Average Weekday Subway Ridership data and fifteen traffic locations from the 2014-2018 Traffic Volume Counts data. The difference in daily ridership among the ten stations is 8%, while the median difference in the traffic volumes among the traffic sites is 29% (Chow et al., 2020).

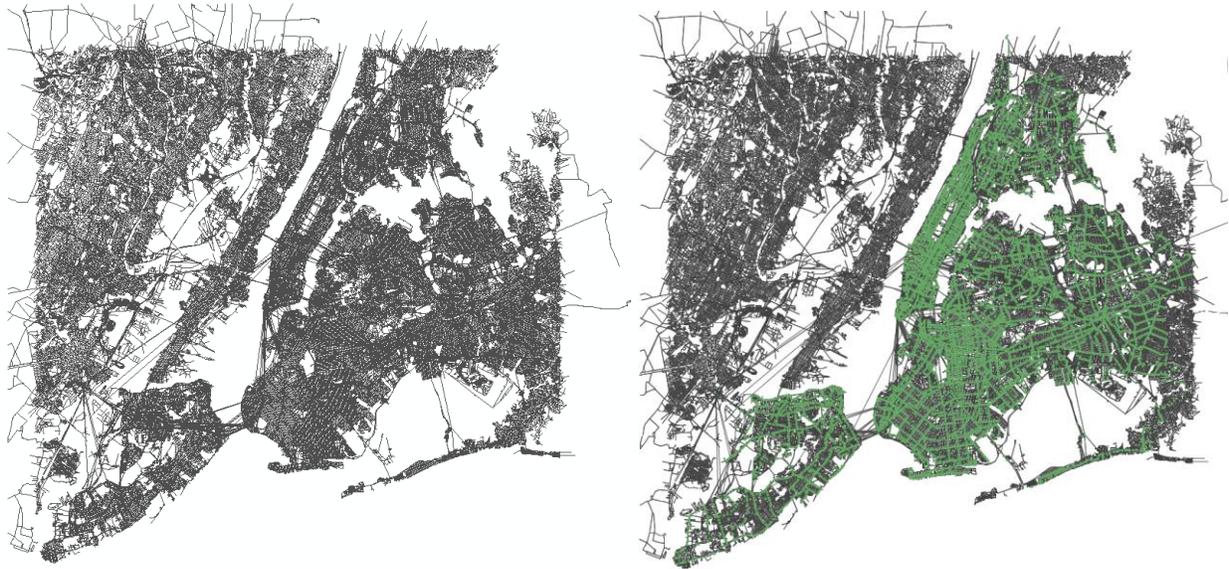

Figure 3. Input road network (left) and transit network (right) of NYC

## 2.3. Scenarios

This is not the first time that a public health crisis that has widely impact the traffic system, especially put the transit system in turmoil. The SARS epidemic in 2003 that first spread in China caused more than 50% of transit ridership reduction in major cities such as Beijing, Shanghai, and Hong Kong. It took about a year for urban transit ridership to recover after the epidemic. In addition, as many cities gradually reopen after the COVID-19 pandemic, there are many strategies and lessons that we can learn during the outbreak and through recovery.

The public transit agency in Beijing, China, published their two-phase reopening plan (Pei, 2020). In phase one, businesses reopen but schools remain closed. The predicted commute traffic returns to normal, but decreased secondary traffic (i.e. shopping, leisure activity) leads to a maximum transit ridership that is only 50% of the normal condition. In phase two, all the businesses and schools reopen, with all commute traffic returning to normal, but still with decreased secondary traffic, leading to a predicted maximum ridership of 75% of the normal setting. The data shows a 20% and 40% restoration of subway ridership one and two months after reopening in Beijing. In Shanghai and Guangzhou, an average rate of 40% and 63% of ridership were restored after 1 month and 2 months of reopening, compared to the pre-COVID period, respectively (Gao et al., 2020).

A mode shift towards walking and cycling was observed after the city reopened in China. According to a survey conducted by the Institute for Transportation and Development Policy (ITDP, 2020) in early March 2020 in Guangzhou, China, 40% of previous transit commuters shifted to private cars, taxis, and ride-hailing, with more to walking and biking. The bike share systems in Beijing (China) saw an increase in usage by 150% from February 10 to March 4, 2020, one month after reopening. Continued monitoring of the mode shift and repositioning of transit services with new health standards can provide insight during the anticipated upcoming recovery period in NYC.

In many cities, transit ridership has rebounded during reopening, but is still lower than before the pandemic. Ridership on the Tokyo Metro dropped by 60% in April 2020 during the lockdown period and reached 63% of normal levels by August after reopening (Sam Schwartz, 2020). Ridership on the Paris metro system dropped to just 5% of pre-pandemic rates by April 2020 and by late June 2020 recovered to 55% of pre-pandemic rates (Sam Schwartz, 2020). Transit ridership remains far below pre-COVID levels, but car traffic has roared back. Even with more people working from home after the outbreak of COVID-19, INRIX has reported that VMT across the United States surpassed 100% of normalized VMT in July 2020 (Markezich, 2020).

To ensure social distancing, some cities have reduced vehicle capacity for public transit during the shutdown and reopening periods. For example, the Metropolitan Transit Authority of Harris County (METRO), Texas, reduced seating by 50% by tagging seats as unavailable during the shutdown and the reopening as of July 2020. When buses reach capacity, digital signs advise individuals to wait for the next bus (Houston Metro, 2020). New Jersey Transit Corporation (NJ TRANSIT) was ordered to cut capacity to 50% until July 15, 2020, when the state entered reopening stage 2, after which full-capacity operations resumed (NJ TRANSIT, 2020). Despite the losses in efficiency, this can be a feasible solution to reduce contact risk and encourage people to use transit.

To study the different strategies under the reopening scenarios, the following scenarios (10 scenarios in total) are examined:
- **Pre-COVID-19 scenario (Jan 2020)**: we use the calibrated MATSim-NYC model updated with January 2020 transit schedules.
- **During COVID-19 scenario (Mar 2020)**: we use the MATSim-NYC-COVID model that was recalibrated using the mobility data during COVID stay-at-home period in NYC.
- **Reopening Phase 1 – 4 without transit capacity restrictions (4 scenarios)**: the recalibrated MATSim-NYC-COVID model is used with modifications to the WFH proportions according to the NY State plan. No transit capacity restriction is applied.
- **Reopening Phase 1 – 4 with transit capacity restrictions (4 scenarios):** a lower transit occupancy is enforced in these scenarios. A 50% transit capacity restriction is applied to forecast both the impact on transit ridership as well as the road traffic.

## 3. Proposed MATSim-NYC-COVID model

### 3.1 Agenda recalibration
The COVID model requires recalibrating the mode choice model to fit the new behavioral setting. The new model is based on COVID WFH agendas, and the mode choice models are then recalibrated to fit the trip reduction data in Table 1.

Determining the work-from-home rate is crucial in the COVID model development. Using NAICS codes consistent with the employment industries in the synthetic population, the work-from-home rates for all occupations can be computed (Dingel and Neiman, 2020). Table 3 shows the classification of teleworkable employment for different industries resulting in the proportions. The reopening proportions are obtained from New York State and discussed further in Section 4.

A new synthetic population was developed based on randomly re-assigned individuals to WFH based on the proportions for each industry. A 44% overall WFH rate was estimated for this newly synthesized population, which is very close to the result (42%) from Dingel and Neiman (2020). In addition, all school/university and secondary trips are assumed to remain closed in this period. Figure 4 shows the total number of workers who are not work-from-home for each traffic analysis zone (TAZ) in the Pre-COVID setting (left) and the COVID setting (right).

Table 3. Non-WFH rates during COVID stay-at-home and reopening phases

| ID | Industry | Non-WFH proportions | | | | |
|---|---|---|---|---|---|---|
| | | COVID | Phase1 | Phase2 | Phase3 | Phase4 |
| 1 | not working | 0 | 0 | 0 | 0 | 1 |
| 2 | Agriculture, forestry, fishing and hunting, and mining | 0.92 | 1 | 1 | 1 | 1 |
| 3 | Construction | 0.81 | 1 | 1 | 1 | 1 |
| 4 | Manufacturing | 0.78 | 1 | 1 | 1 | 1 |
| 5 | Wholesale trade | 0.48 | 1 | 1 | 1 | 1 |
| 6 | Retail trade | 0.86 | 0.93 | 1 | 1 | 1 |
| 7 | Transportation and warehousing, and utilities | 0.72 | 0.72 | 0.72 | 1 | 1 |
| 8 | Information | 0.28 | 0.28 | 0.28 | 0.28 | 1 |
| 9 | Finance and insurance, and real estate and rental and leasing | 0.41 | 0.41 | 1 | 1 | 1 |
| 11 | Professional, Scientific, and Technical Services | 0.2 | 0.2 | 1 | 1 | 1 |
| 12 | Management of Companies and Enterprises | 0.21 | 0.21 | 0.605 | 1 | 1 |
| 13 | Administrative and Support and Waste Management and Remediation Services | 0.69 | 0.69 | 0.69 | 1 | 1 |
| 14 | Educational services, and health care and social assistance | 0.46 | 0.73 | 1 | 1 | 1 |
| 15 | Arts, entertainment, and recreation, and accommodation and food services | 0.83 | 0.83 | 0.83 | 0.915 | 1 |
| 16 | Other services, except public administration | 0.69 | 0.69 | 1 | 1 | 1 |
| 17 | Public administration | 0.59 | 0.59 | 1 | 1 | 1 |

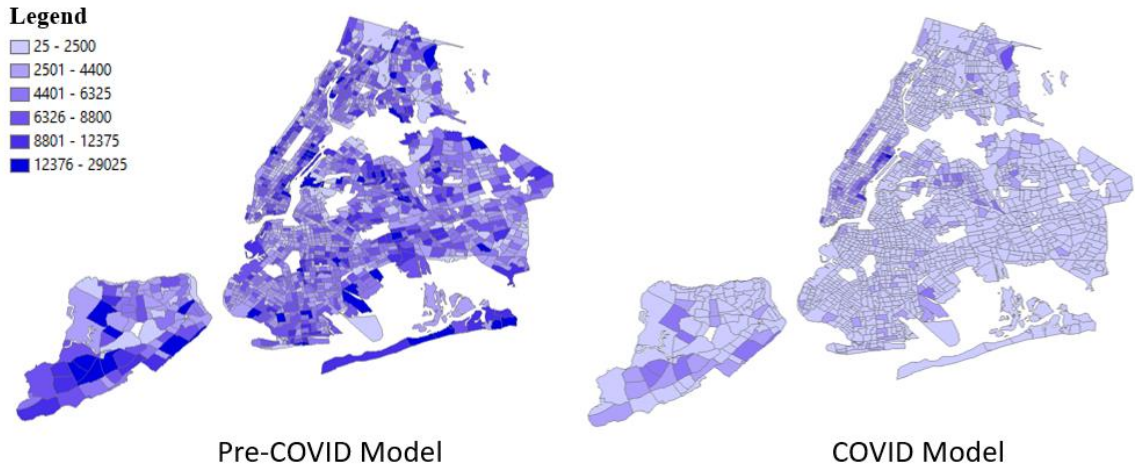

Figure 4. The number of outside-home workers by TAZ of residence in Pre-COVID model (left) and COVID model (right).

The road network in COVID model is kept the same as the Pre-COVID model. The transit schedule during COVID period is generated from the GTFS data from March 20, 2020.

### 3.2 Mode choice model calibration
With the newly recalibrated population that accounts for WFH, we recalibrated the mode choice utility functions based on the ridership reduction data shown in Table 1. Four mode parameters were updated and calibrated, namely the ASC values for transit, driving, walking, and biking. As shown in Table 2, those parameters were originally $2.95, -0.06, 5.73$, and $0.44$, for Manhattan population and $0.76, -0.05, 3.49$, and $-1.35$ for the non-Manhattan population, respectively.

The calibration approach is based on the SPSA algorithm (Spall, 1988, 1998a, 199b), which uses random perturbations from a current point to measure the gradient across multiple dimensions cheaply. The procedure can be summarized as following: with the input data and initial parameters, we run the simulation and compare the results with observed data and calculate the predefined loss function as the simulation error. If the resulting simulation error is found to be acceptable, we terminate the calibration process. Otherwise, the parameters are updated using SPSA algorithm and then re-run the simulation until the accuracy is satisfied. A similar calibration method for simulation models can be found in our MATSim-NYC model (Chow et al. (2020)). In this study, the stopping criterion is set as an average absolute difference in trip reduction lower than 0.1, including subway, car, and walking, which converged after 15 iterations.

The new ASCs under COVID are shown in Table 4. The first number refers to the updated ASC (e.g. 1.95 for transit for Manhattan) while the second number in parenthesis refers to the absolute change from the original parameter (e.g. -1.00 is a reduction from an original of 2.95 for transit for Manhattan).

The ASCs of the COVID model show that the mode preference for driving increased dramatically compared to pre-COVID setting while the preference for public transit decreased. In the Pre-COVID model, travelers in Manhattan have a much higher preference of using transit than other regions, but the decrease of transit preference in Manhattan is more significant than other boroughs in the COVID model. The mode preference of car increased more in the Non-Manhattan boroughs compared to Manhattan, likely because Manhattan has lower car ownership compared to other boroughs.

Table 4. Changes in mode choice utility function alternative-specific constants due to COVID

|  | **Manhattan ASC (+/-)** | **Non-Manhattan (+/-)** |
| ---: | :---: | :---: |
| **Transit** | 1.95 (-1.00) | 0.36 (-0.40) |
| **Driving** | 3.07 (+3.13) | 3.56 (+3.61) |
| **Walking** | 8.53 (+2.80) | 6.00 (+2.51) |
| **Biking** | 1.94 (+1.50) | 0.00 (+1.35) |

Table 5 shows the comparison of the ridership reduction and the reference data (Table 1) along with the percentage point difference (p.p. difference). The simulation results show that trip reductions during the pandemic for subway, car, and walking are 91%, 76%, and 68%, respectively, compared to observed reductions of 89%, 58%, and 76%. The differences are likely from a combination of different factors. MATSim-NYC was built assuming bus and subway have the same utility parameters, and there are still secondary trips being made during the stay at home order, many of which are probably by car (to purchase home supplies and groceries, for example).

Table 5. Compare the trip reduction in the simulation with real data

| **Mode** | **Simulated Reduction** | **Real Reduction** | **p.p. difference** |
| :---: | :---: | :---: | :---: |
| **Subway** | 91% | 89% | 2% |
| **Car** | 76% | 58% | 18% |
| **Walk** | 68% | 76% | -7% |

### 3.3 Validation of the MATSIM-NYC-COVID model

To validate the simulation, the subway station entrance data from MTA is used to compare with the results from our baseline models. Table 6 shows a comparison of some of the most popular stations in NYC. The nine busiest stations with reliable data during COVID-19 are selected out of 472 stations, where the ridership among this subset makes up nearly 15% of the total daily subway ridership in NYC as a sufficient sample. The MTA subway daily ridership data in 2019 is used to compare with the Pre-COVID model. The daily average entrance data was collected from MTA turnstile dataset during March 23, 2020 to April 12, 2020.

Most of the stations have close predictions to the observations. The higher deviations in ridership of 34 St - Herald Square, 42 St - Bryant Park, Atlantic Av-Barclays Ctr might be due to passengers arriving at nearby railway stations (e.g., LIRR, Penn Station) or bus terminal (e.g., Port Authority Bus Terminal). The deviations in Jackson Hts-Roosevelt Av in the COVID model might be related to less passengers heading to LaGuardia Airport, and 59 ST - Columbus Circle in the COVID model might be related to the tourist trips. The current model does not incorporate non-resident trips, tourist trips and airport-related trips due to the limitations of the data. Spatial heterogeneity of transit preference may also affect the prediction accuracy. Some areas might have a higher proportion of people who must take transit and result in a lower reduction in ridership. However, we only calibrated the alternative specific constant for the transit mode, which on its own cannot capture this spatial heterogeneity of transit preference. The average difference in ridership among those selected stations is 5.67% in the Pre-COVID model and 7.91% in the COVID model, whereas the mean absolute percentage difference is 18.54% in the Pre-COVID model and 20.93% in the COVID model. These results are acceptable for citywide planning models in practice (see Flyvbjerg et al., 2005).

Table 6. Comparison of the real and simulated station entrance data

| Station | Pre-COVID Model | | | COVID Model | | |
|---|---|---|---|---|---|---|
| | Simulated | Real | diff% | Simulated | Real | diff% |
| Times Square - 42 St | 191,425 | 202,363 | -5.4% | 5,675 | 5,033 | 12.8% |
| 34 St - Herald Square | 124,500 | 125,682 | -0.9% | 7,400 | 5,610 | 31.9% |
| 14 St - Union Square | 97,825 | 106,718 | -8.3% | 6,900 | 6,566 | 5.1% |
| 59 ST - Columbus Circle | 75,050 | 73,836 | 1.6% | 5,925 | 4,448 | 33.2% |
| Flushing - Main St | 56,503 | 72,475 | -22.0% | 5,175 | 5,596 | -7.5% |
| 47-50 Streets - Rockefeller Ctr | 63,609 | 53,700 | 18.5% | 1,850 | 2,451 | -24.5% |
| 42 St - Bryant Park | 58,339 | 38,925 | 49.9% | 7,250 | 8,342 | -13.1% |
| Jackson Heights - Roosevelt Av (Queens) | 41,200 | 52,296 | -21.2% | 1,550 | 1,056 | 46.8% |
| Atlantic Av - Barclays Ctr (Brooklyn) | 59,350 | 42,711 | 39.0% | 4,600 | 5,315 | -13.4% |
| Average diff% | | | 5.67% | | | 7.91% |

### 3.4 Comparing the result of Pre-COVID model and COVID model

To further study the impact of COVID-19 on the traffic system, we compare the simulation outputs between the Pre-COVID and COVID models. The result shows that COVID-19 and social distancing practices have changed people's travel behavior and may reshape the mode share (Figure 5). Compared to the pre-COVID period, the mode share of transit decreased about 19%, while the mode share of cars increased about 6% in the COVID model. The total number of trips also decreased dramatically in the COVID model, so an increase in mode share of car does not mean that the total number of car trips increased.

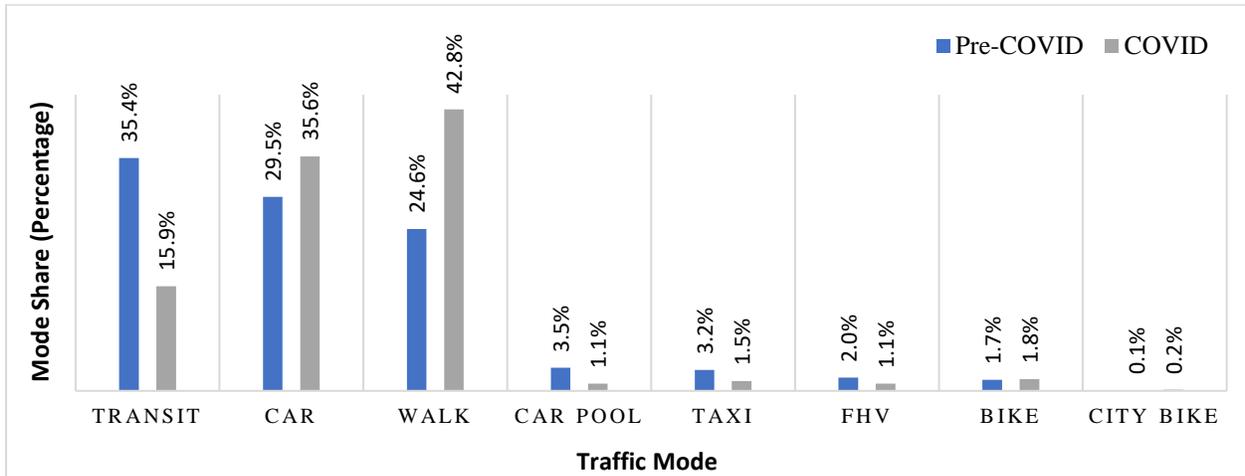

Figure 5. Compare the mode share in Pre-COVID model and COVID model.

Figure 6 shows the spatial distribution of link speed in the Pre-COVID and COVID model using the MATSim visulation tool, Simunto Via. Clearly, with the "stay-at-home" order, the volume of vehicles on the roads have dropped leading to increased travel speeds. These numbers are readily confirmed by the traffic monitoring reported in the C2SMART White Paper on COVID-19 Impacts Issue No. 2 (Gao et al., 2020).

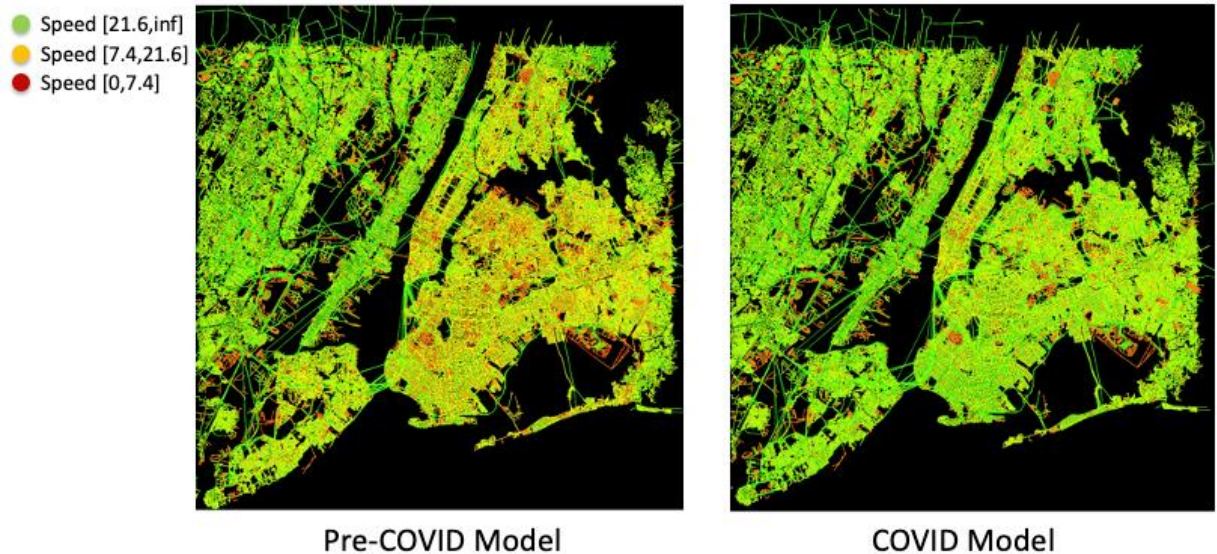

Figure 6. The link speed comparison at 9am in the morning peak (km/h)

**3.5 Sensitivity analysis of stay-at-home restrictions on employment industries**
Having calibrated and validated the model, we seek to better understand the sensitivity of road traffic and transit on different employment industries. According to the 2015 American Community Survey, only 44% of households in NYC own cars. Nonetheless, a study by INRIX in 2019 found that NYC ranks as the fourth congested city in the U.S. costing about $11B a year because of time lost in traffic (Reed, 2019). This could be even worse because of COVID-19.

It is possible for NYC to propose an employer-based restriction on employees to have a portion of them continue working from home, or to rotate employees in shifts during the reopening phases. Data shows that remote work has increased by 44% from 2015 to 2020, and remote work opportunities are more common in high income areas because they are more likely tied to information-based positions that can be performed off-site (Griffith, 2020). However, it remains unclear which industries would have the greatest impact if they were able to adopt such policies.

We investigate the relationship between industry and traffic mode choice using our agent-based simulation model. Figure 7 shows the changes in number of trips by car (a) and transit (b) in the Pre-COVID model and COVID model. The result shows that although many employees are working at home, the car trips in some industries in the COVID model became higher than the Pre-COVID model, likely due to the mode shift to driving. For example, the combination of home/work locations and mode preferences for industry #6 (Retail Trade) and #15 (Arts) led to more car trips during COVID. In addition, different industries have different levels of transit trip reduction. These findings also provide suggestions to policymakers on which industries to target for demand management strategies. If policymakers were to focus on specific industries to reduce car trips during a reduced transit capacity reopening, they might want to consider high auto mode usage industries like Retail Trade and Administrative industries (#6 and #13).

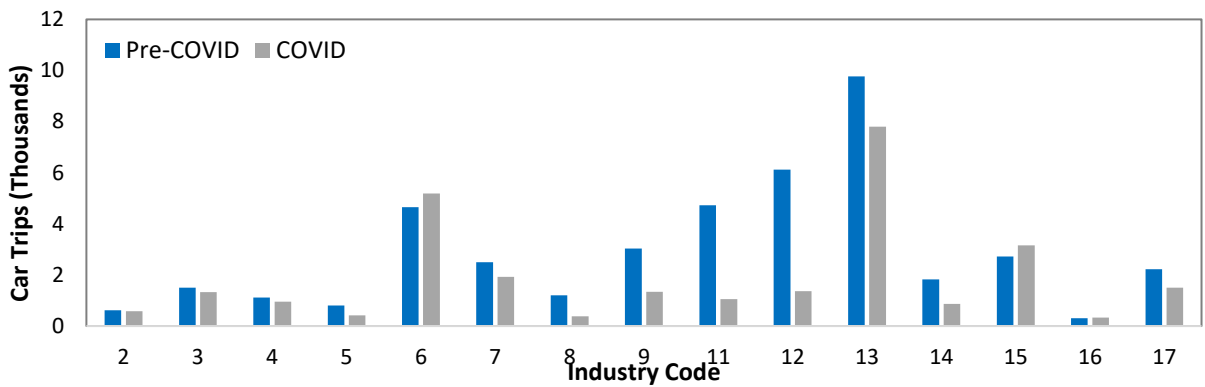
(a) Car trips

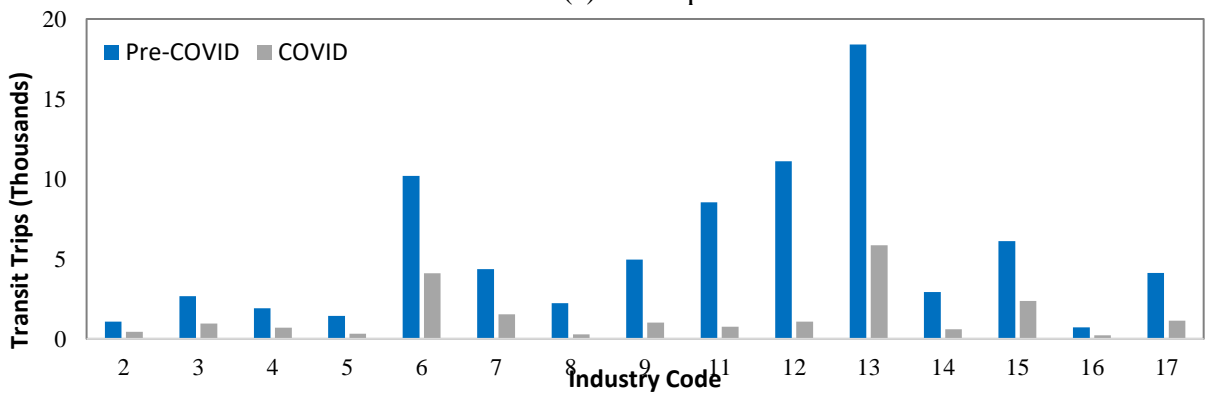
(b) Transit trips

Figure 7. Number of trips changes of (a) car and (b) transit, in Pre-COVID and COVID models.

# 4. Analysis of reopening plan under different transit operating strategies

New York State is planning a four-phase reopening based on the regional guidelines for reopening NYC, according to the Regional Guidelines on May 8, 2020 (New York State, 2020). The 'Un-Pause NY' approach is designed to open businesses in phases of priority. Businesses considered "more essential" with inherent low risks of infection in the workplace and to customers are prioritized, followed by businesses considered "less essential" or those that present a higher risk of infection spread. Based on the priority industries in each phase, we generate synthetic populations corresponding to those employees returning to work. Table 3 shows the percentage of employees who will return to work in different industries each phase.

A study from China shows that even after a city is re-opened, many people may still think driving car is safer than using transit (ITDP, 2020). There could also be a boom in other traffic modes, such as walking and cycling. In the reopening scenarios, a key assumption here is that the mode preference during the reopening phases mirrors what was observed during the pandemic period due to behavioral inertia. In other words, the mode choice model parameters in the reopening scenarios are assumed to be the same as in the COVID model. This assumption can be conservative and should be treated as worst-case scenarios for transit. The current model will be continuously enhanced when more data becomes available. For the first two reopening phases, the transit schedule was assumed to be the same as it was during the COVID period (using the GTFS on Mar 20, 2020). When simulating the last two reopening phases, the regular transit schedule (GTFS on Jan 20, 2020) is applied. The simulation model for each scenario is run for 100 iterations and the computation time for each scenario is on average of about 7.5 hours using an Intel Xeon 2.1 GHz with 64 GB RAM.

## 4.1. Scenario Set 1: No transit capacity restriction

In these scenarios, there is no transit capacity restriction while assuming mode preferences held during the crisis are maintained. Figure 8(a) shows the trip ratio in the COVID model and in the four reopen phases in Scenario Set 1 compared to the Pre-COVID model. The result shows that only 73% transit ridership will be back by Phase 4 when the city full reopens. The number of car trips increases 1.42 times compared with the number of pre-COVID trips. In addition, the number of walking and biking trips increases in Phase 4 as well. Figure 9 shows the changes of mode share in each phase. The transit mode share in Phase 4 in this scenario set decreases by 9.5 percentage points compared to the pre-COVID period while the mode share of car increases 12.3 percentage points. There are relatively small changes in mode share in other traffic modes because their total number of trips are low compared to transit and car trips.

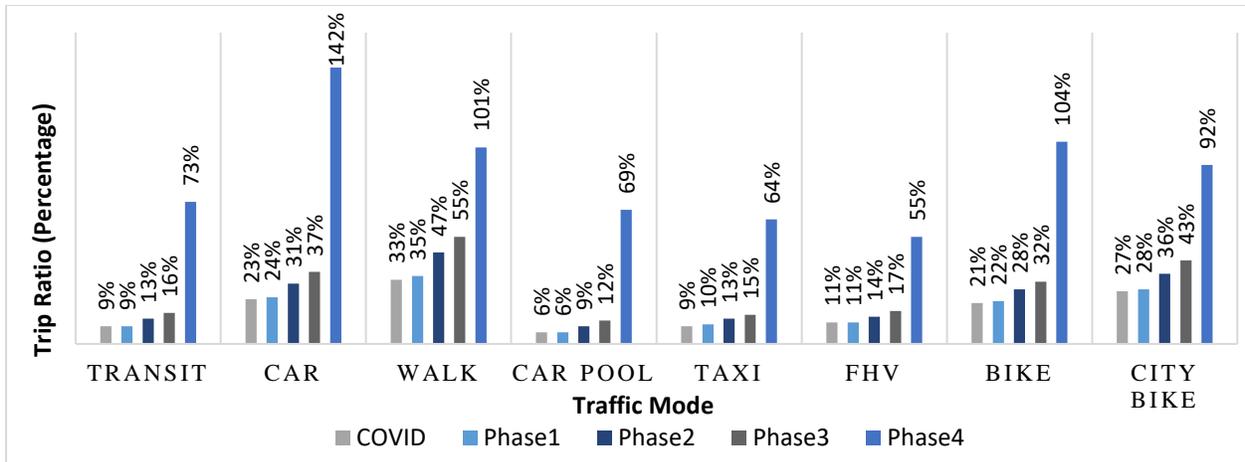

(a) Scenario set 1

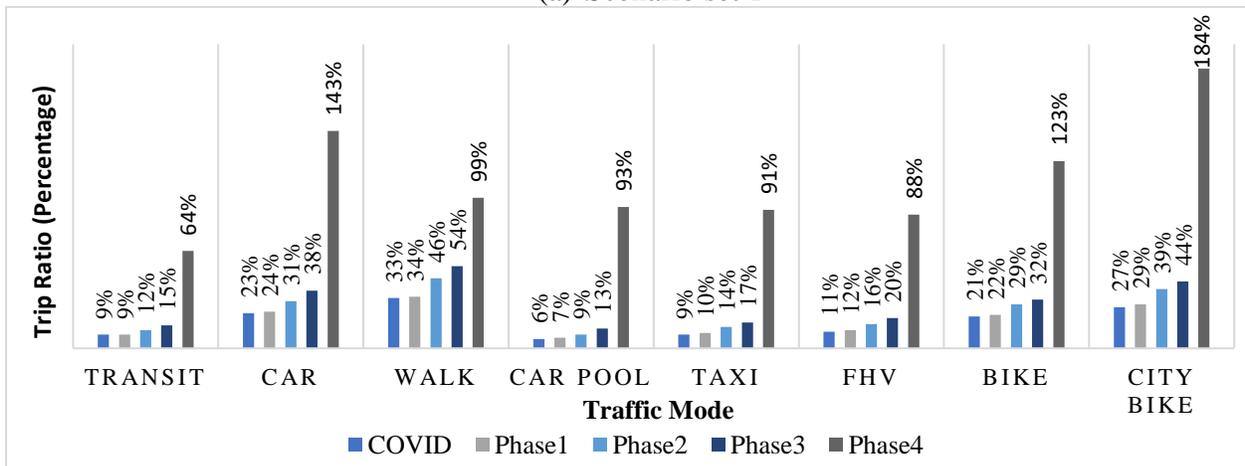

(b) Scenario set 2

Figure 8. The trip ratio in the COVID model and reopen phases in two scenario sets compared to the Pre-COVID model.

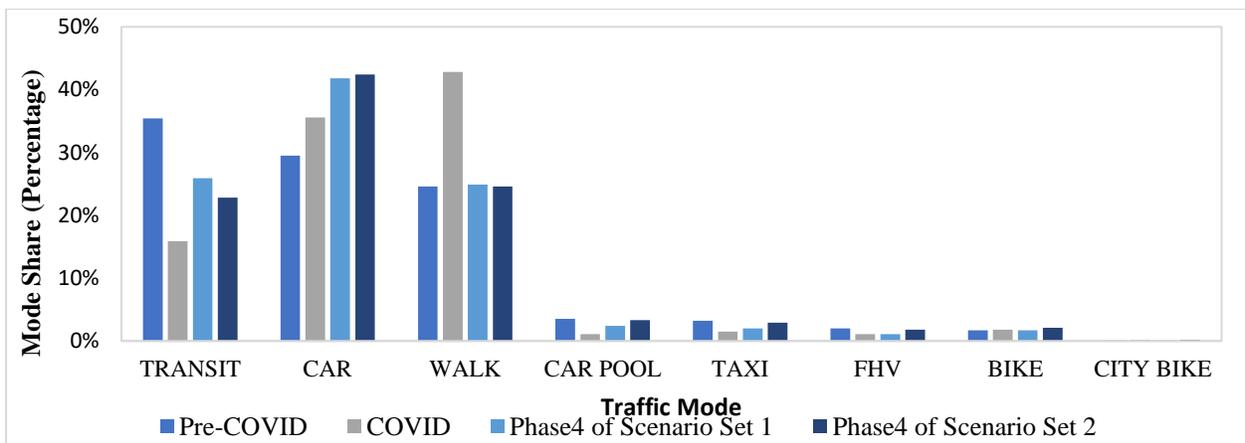

Figure 9. The changes of mode share in the Pre-COVID model, COVID model and Phase 4 in two simulation scenario sets (with and without transit capacity restriction).

This result confirms the concern shared by many transportation professionals that the number of the transit trips may not return to original state even without transit capacity reduction due to behavioral inertia, resulting in a significant increase in car trips (42% increase compared to a 27% decrease in transit trips).

**4.2. Scenario Set 2: 50% Transit capacity restriction**
To ensure the social distance, we assume a 50% transit capacity restriction on all transit vehicles in the second set of scenarios, assuming mode preferences held during the crisis are maintained. Figure 8(b) shows the trip ratio in the COVID model and the reopen phases in scenario set 2 compared to the Pre-COVID model. After the city fully reopens in phase 4, 64% transit ridership is projected with a 143% increase in car trips in Scenario Set 2. Compared with the results in Scenario Set 1, it suggests that the strategy to restrict transit capacity to 50% on its own would not be the primary contributor to additional car traffic. Nonetheless, it can still lead to a further reduction in transit ridership. It is also interesting to see that bike and Citi Bike modes both grow substantially when transit capacity is restricted, which suggests policymakers should plan accordingly with those modes, and perhaps with scooters as well, in meeting the shift in demand from transit.

The changes in mode share under reopening Phase 4 in Scenario Set 2 compared the Pre-COVID period and COVID period are shown in Figure 9. The mode share of transit in phase 4 may decrease 12.6 percentage points compared to the Pre-COVID period, which is about 3 percentage points lower compared to Scenario Set 1. The mode share for car may increase 12.9 percentage points compared to the Pre-COVID period and increase 0.6 percentage points compared to Scenario Set 1. There is also an increase in mode share in other traffic modes compared to Scenario Set 1, such as bike, taxi and FHV.

**4.3. Comparing the traffic condition in full reopen phase (phase 4) and in the Pre-COVID model**
To test the impacts of behavior inertia from COVID-19 with different transit operations in the full reopening Phase 4, we estimate the traffic volume and speed for car trips per road link from the simulation results.

Table 7 shows the changes of travel time and distance per car trip in the Phase 4 of reopening compared to the Pre-COVID model. "Manhattan" in the table refers to trips with destination in Manhattan while "Citywide" indicates all trips in the simulation model. The results show that both travel time and distance per car trip in Manhattan increased. For example, the number of car trips is 2.57 times of the Pre-COVID period under no transit capacity restriction and 2.77 times of the Pre-COVID period under 50% transit capacity restriction. The results also indicate a much higher increase in travel time and distance with 50% transit capacity restriction compared to no capacity restriction.

The citywide metrics show an increase in total number of trips, total travel distance and total travel time, while a small decrease in average travel distance and average travel time. This might be because many increased car trips are short-distance trips can lead to a decrease in average travel time and distance.

Table 7 Travel time and distance per car trip in Phase 4 compared to the Pre-COVID model (percentage increase)

| Scenario/Region | Total Car Trips | Avg Distance | Sum Distance | Avg Travel Time | Sum Travel Time |
|---|---|---|---|---|---|
| **Scenario Set 1 (Manhattan)** | +157% | +2% | +162% | +29% | +232% |
| **Scenario Set 2 (Manhattan)** | +177% | +7% | +197% | +60% | +342% |
| **Scenario Set 1 (Citywide)** | +42% | -3% | +38% | -9% | +30% |
| **Scenario Set 2 (Citywide)** | +43% | -1% | +42% | -4% | +37% |

Figure 10 shows the average link speed and volume in the Pre-COVID model and Phase 4 in the two reopening scenario sets to better understand the traffic congestion in the road network. The result indicates the impact of behavior inertia after the city reopens in Phase 4 will increase the average link volume and decrease the average link speed, especially during the peak hour, and the result is significant compared to the Pre-COVID mode. In addition, the average changes in Manhattan are higher compared to the average value of all regions, but the difference between Scenario Set 1 and Scenario Set 2 is relatively small.

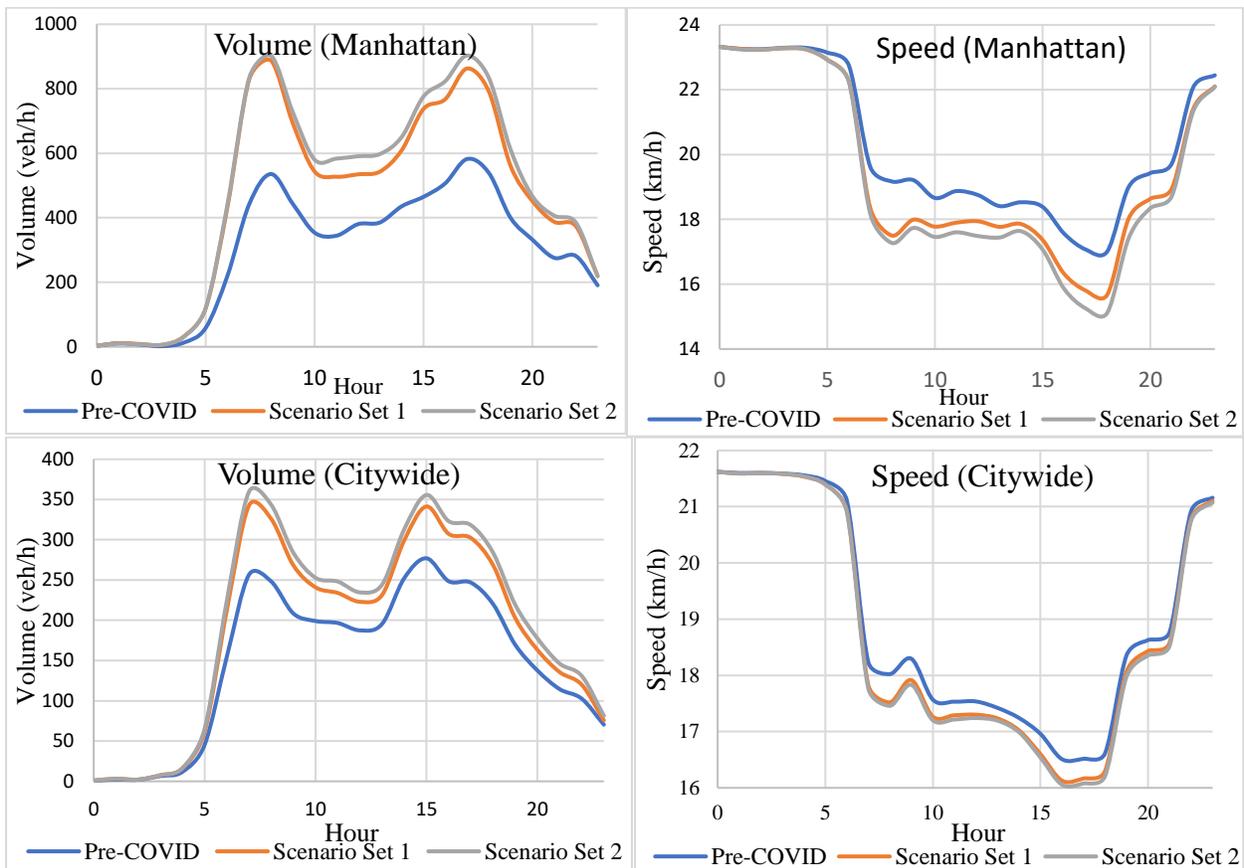

Figure 10 Average link volume and speed by region in the Pre-COVID model and Phase 4 in two reopening scenario sets

**4.4. Comparing the average change in consumer surplus of agents**
In MATSim, each agent has a daily plan of trips and activities. The plan of agents is executed in the simulation and each plan is then scored and assigned with a utility. In general, the score includes the travel disutility from the chosen mode plus the disutility of any schedule delay past the desired arrival time and reflects the consumer surplus. MATSim includes a score for duration of an activity conducted, up to a maximum duration, but we leave that portion out since it is not calibrated. The goal of each agent is to maximize the utility of plan by re-planning based on a co-evolutionary algorithm (see Horni et al., 2016). When the parameters of the score functions are unchanged, comparisons can be made between scenarios. We make such a comparison between the scores, which are converted to monetary values ($) using the $29/h value of time from the mode choice model for Manhattan trips (He et al., 2020a; Chow et al., 2020).

Table 8 shows the change in average consumer surplus of all the agents in Scenario Set 2 relative to Scenario Set 1. The table indicates that the transit capacity reduction would have an adverse effect overall on consumer surplus. Most of that effect is on the car mode, which is likely because of the increase in car trips under a much more congested setting (from 1.42 times to 1.43 times the pre-COVID trips). The impact on other modes in Manhattan is relatively minor in comparison, which suggests that there is adequate supply in the other modes in Manhattan to absorb the shift in ridership to them. The scores offer a different picture of the transit capacity reduction; while it seems like the additional car traffic would be minor compared to Phase 4 without any transit capacity reduction, they have an impact on the already heavily congested traffic network. Implementing such a strategy should not only look at improving micromobility capacity but also toward congestion alleviation strategies, particularly in Manhattan where there is supply from other modes to absorb the shift in trips.

Table 8. Average change in daily consumer surplus in Scenario Set 2 relative to Scenario Set 1

|  | Citywide | | Manhattan | |
| --- | --- | --- | --- | --- |
|  | All modes | Car | All modes | Car |
| **Scenario Set 2** | -$36.15 | -$61.84 | -$1.59 | -$57.89 |

# 5. Conclusion

Although NYC entered Phase 1 of reopening from COVID-19 stay-at-home orders that persisted since mid-March 2020, many new challenges for transportation systems remain. There has not yet been a concrete solution, especially for mass transit, to increase service while protecting riders on crowded subway or buses. The simulation models proposed in this study are designed to evaluate the impact of COVID-19 on mass transit ridership as well as helping policy-makers plan for post-coronavirus. Among various insights gained from the study, the following key points are made:
- There is a clear behavioral change due to COVID shifting away from shared use modes to driving, walking, and biking, and this change differs between Manhattan residents and non-Manhattan residents.
- The behavioral change results in some employment industries taking more car trips to work; policymakers wishing to complement transit reduction policy with employer-enforced stay-at-home rotations/curfews/shifts may consider targeting certain high car-trip segments like Retail Trade and Administrative.

- If policymakers wish to instead pursue outreach to encourage certain population segments to revert their behavior back, then the key industries are Retail Trade and Arts where car traffic in COVID has presumably increased over pre-COVID condition.
- Due to the behavioral inertia, a reopening that does not restrict transit capacity would still only operate at 73% ridership from pre-COVID while increasing car traffic to 142% level.
- Transit capacity restriction, even by 50%, would reduce transit ridership from 73% to 64% of pre-COVID ridership. Interestingly, car trips would only marginally increase further from 142% to 143%. It seems the transit capacity restriction would shift most users to non-car modes, e.g. bike and Citi Bike. The impacts of these shifts are disproportionate, however. The small increase in added car trips to an already congested traffic setting exacerbates the loss in consumer surplus for drivers. For other modes, it is not as significant of a loss, particularly in Manhattan (suggesting adequate redundant supply to capture the shift in trips). The findings are generally encouraging for the policy, but also suggests that policymakers should plan accordingly to accommodate that increase in active transportation, perhaps even considering the use of other forms of micromobility like e-scooters, and the increased cost in traffic congestion with Manhattan-oriented alleviation strategies (cordon-based pricing may help).

The open-source, modular nature of the developed simulation model in MATSim can be applied to evaluate many new strategies and policies for city government and traffic agencies, such as the impact of new bike lanes and different transit operations. Besides the effects of the pandemic and an ensuing recovery on transit use, air quality and emission impact estimation during the COVID-19 pandemic and reopening can also be studied in the future. The current model will also be continuously enhanced when more data becomes available.

There are also some limitations of this study. With limited data, the mode choice multiplier for certain traffic modes were not calibrated in the "stay-at-home" model, which means they are assumed to remain unchanged with respect to each other. The current model will be continuously enhanced when more data becomes available. Moreover, a 4% sample population is used in the current model instead of the entire population due to the large computation time. This can be further improved in the future.

## Acknowledgments

This research was conducted with the support of the C2SMART University Transportation Center.